\documentclass[
    ,final            
  ]
  {aipproc}

\layoutstyle{6x9}
\usepackage{graphics}
\usepackage{graphicx}
\usepackage{amssymb}
\usepackage[rightcaption]{sidecap}
\usepackage{subfigure}
\usepackage{color}
\definecolor{ah1col}{rgb}{0.84, 0.84, 0.84}

\begin{document}

\title{Differential cross sections and spin density matrix elements for $\gamma p \to \phi p$ from CLAS}


\author{Biplab Dey}{
  address={Department of Physics, Carnegie Mellon University,
  Pittsburgh, PA 15213, USA}
}

\author{Curtis A. Meyer \\(representing the CLAS Collaboration)}{
  address={Department of Physics, Carnegie Mellon University,
  Pittsburgh, PA 15213, USA}
}

\begin{abstract}
Preliminary differential cross-sections and the $\rho^0_{MM'}$ spin density matrix elements (SDME) for the reaction $\gamma p \to \phi p$ for both charged- ($\phi \to K^+ K^-$) and neutral-mode ($\phi \to K^0_L K^0_S$) topologies obtained from CLAS are presented. Our kinematic coverage is from near production threshold ($\sqrt{s} \sim 1.97$~GeV) to $\sqrt{s} = 2.84$~GeV, with a wide coverage in the production angle. As seen in previous LEPS results, the differential cross-sections show a localized ``bump'' between $\sqrt{s} \sim 2$ and 2.2~GeV that is not expected from a simple Pomeron exchange picture. Comparisons between the charged- and neutral-mode results and possible effects from the $K^+ \Lambda(1520)$ channel are discussed. Our SDME results confirm the well-known deviations from t-channel helicity conservation (TCHC) for Pomeron exchange, but s-channel helicity conservation (SCHC) is also seen to be broken.
\end{abstract}

\maketitle


Vector meson electro- and photoproduction have played an important role in our understanding of photon-hadron interactions in QCD. In the so-called vector meson dominance (VMD) model for photoproduction, a real photon can fluctuate into a virtual vector meson, which subsequently scatters off the target proton. For the $\phi(1020)$, an additional attraction is that since this is an almost pure $s\bar{s}$ state, the OZI rule suppresses quark exchanges during interactions with ordinary nucleonic matter, and is therefore considered to be a ``clean'' system to study gluonic exchanges. Recently, the CLAS Collaboration has published detailed cross-section and the unpolarized $\rho^0_{MM'}$ spin density matrix element (SDME) data for the $\omega$ channel~\cite{williams_omega_prc}. In this article, we report the corresponding preliminary results for the $\phi$ channel.

\section{Experimental setup, event selection, and background removal}

The data for this analysis used a tagged photon beam produced via bremsstrahlung, a liquid hydrogen cryotarget, and the CLAS detector system, during the ``$g11a$'' experimental run-period~\cite{williams_omega_prc,cmu_papers} at Jefferson Lab. The reaction of interest was ${\gamma p \to \phi p}$, where the $\phi$ subsequently decays via  $\phi \to K^+ K^-$ (charged-mode, $49.2\%$) and $\phi \to K^0_S K^0_L$ (neutral-mode, $34\%$). For the charged-mode, the outgoing proton and $K^+$ are detected and the $K^-$ is re-constructed as the total missing momenta. For the neutral-mode, the $K^0_S$ is reconstructed from its decay into a $(\pi^+, \pi^-)$ pair, and the $K^0_L$ is reconstructed as the total missing momenta. Therefore the two reaction topologies can be summed up as:
\begin{equation}
\gamma p \to p K^+ (K^-) \;\;(\mbox{charged-mode}) ; \; \gamma p \to p K^0_S (K^0_L) \to p \pi^+ \pi^- (K^0_L) \; (\mbox{neutral-mode}),
\end{equation} 
where the undetected final-state particles are denoted within parentheses. For background removal, a confidence level cut from a kinematic fit to the above reaction topologies, followed by standard timing cuts, were applied. For the neutral-topology, a $0.488~\mbox{GeV} \leq M(\pi^+,\pi^-) \leq 0.508$~GeV cut was placed for $K^0_S$ identification. For the charged-mode, between $\sqrt{s} \sim 2$ and 2.2~GeV, kinematic overlap occurs with the $\gamma p \to K^+ \Lambda(1520) \to K^+ p K^-$ reaction so that a $|M(p,K^-) - 1.520~\mbox{GeV}| \leq 15$~MeV cut was placed for $K^+ \Lambda(1520)$ removal. Next, independent signal-background separation fits were performed in localized regions of phase-space~\cite{williams_omega_prc,cmu_papers} that gave a quality-factor ($Q$-value) for each event, representing the probability of that event being a good signal. To take the finite width of the $\phi$  into account, the signal function was chosen as a Voigtian with a mass-dependent decay width $\Gamma(m)$, while the background shape was kept to be as general as possible. Our final $Q$-value weighted $\phi$ data yields were $\sim$~0.475~million and $\sim$~0.101~million for the charged- and neutral-mode, respectively.

\section{Acceptance Calculation and Differential Cross Sections}

To calculate detector acceptance, 100~million ``raw'' Monte Carlo $\gamma p \to \phi p $ events were generated flat in phase-space and passed through gsim, a GEANT based detector simulator. To incorporate the effect of any physics in the acceptance calculation, next, the production amplitude was expanded in an over-complete basis of $s$-channel $J^P$ waves and an event-based partial wave analysis (PWA) fit was performed employing the extented maximum likelihood method~\cite{williams_omega_prc,cmu_papers}. Based on the results of this PWA fit, a {\em physics-weighted} acceptance was calculated subsequently. 

\begin{figure}
\centering
\hspace{-2.5cm}\includegraphics[angle=90,width=3in]{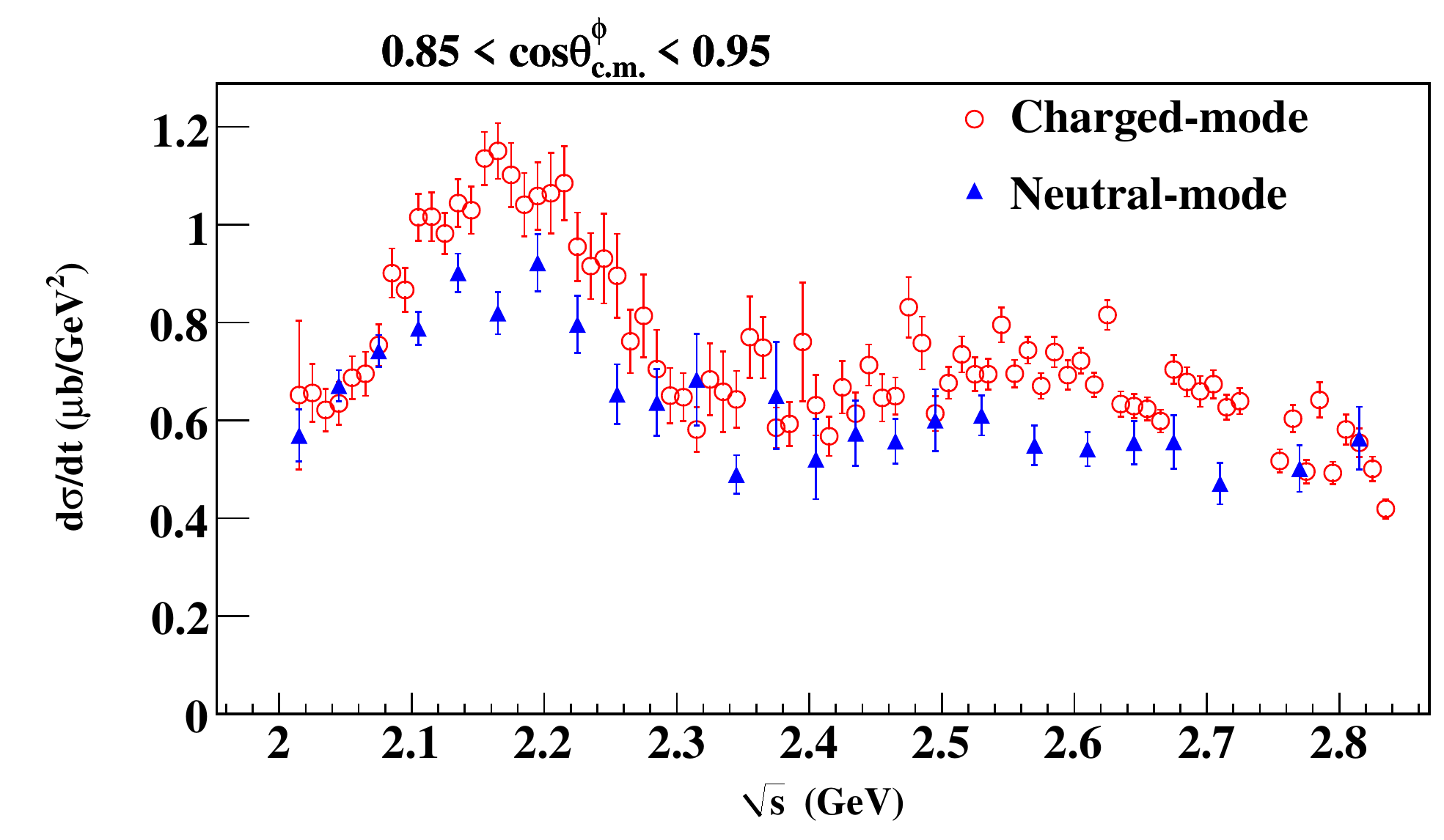}\hspace{-5cm} \vspace{-5cm}\rotatebox{45}{PRELIMINARY}
\caption{\label{fig:dcs} (Color online) Preliminary $\phi$ differential cross sections in a forward-angle bin shown for both the charged- and neutral-mode topologies. The ``bump'' around $\sqrt{s} \sim 2.1$~GeV, seen in previous LEPS data~\cite{mibe} is clearly visible here.}
\end{figure}

\begin{figure}
\centering
\hspace{-2.2cm} \includegraphics[width=2.6in]{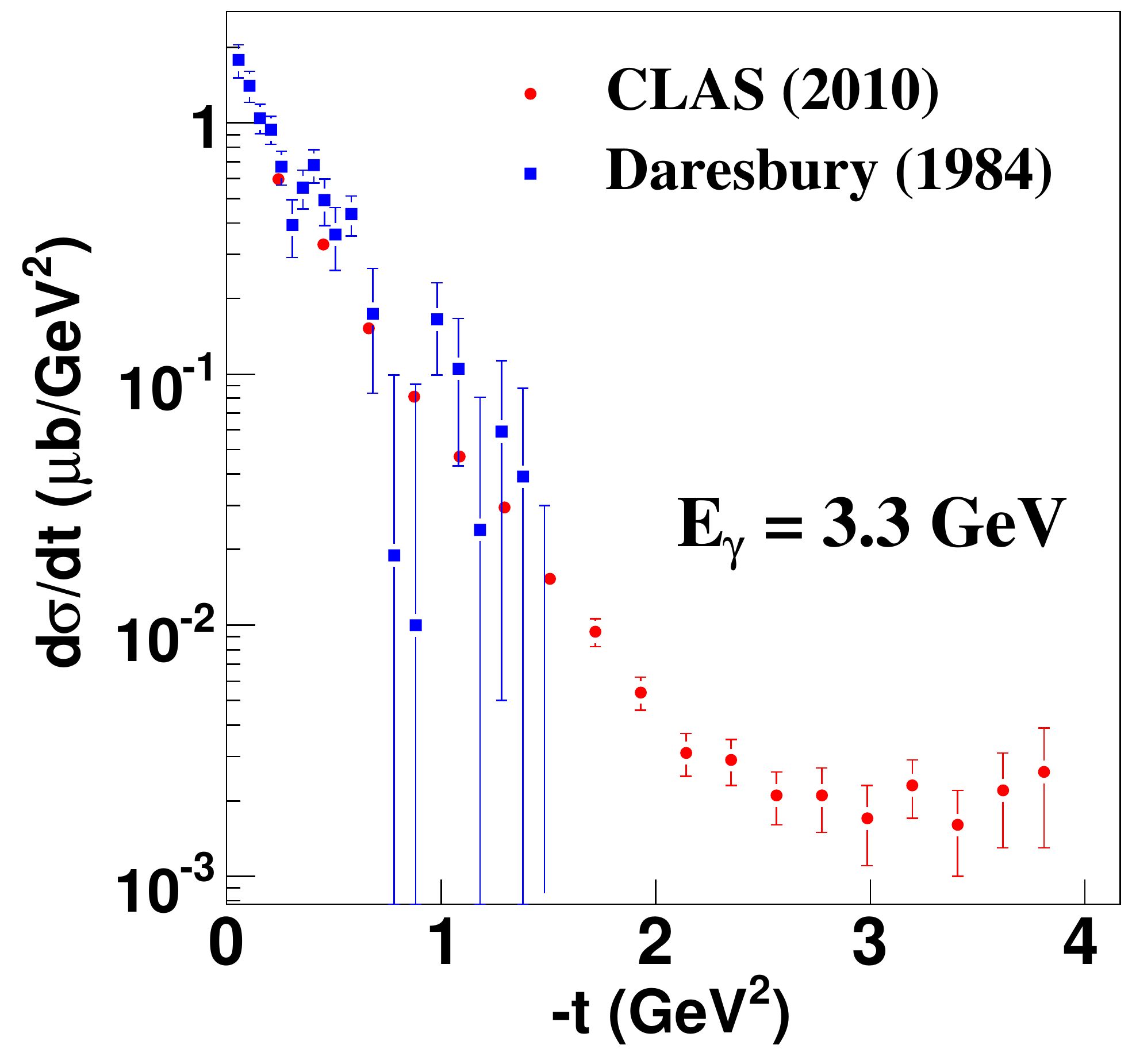}\hspace{-4cm} \vspace{-10cm}\rotatebox{45}{PRELIMINARY}
\caption{\label{fig:barber} (Color online) Comparion between the present CLAS and 1984 Daresbury data~\cite{barber} at $E_\gamma = 3.3$~GeV.}
\end{figure}

\begin{figure}
\centering
\hspace{-2.5cm} \includegraphics[angle=90,width=2.5in]{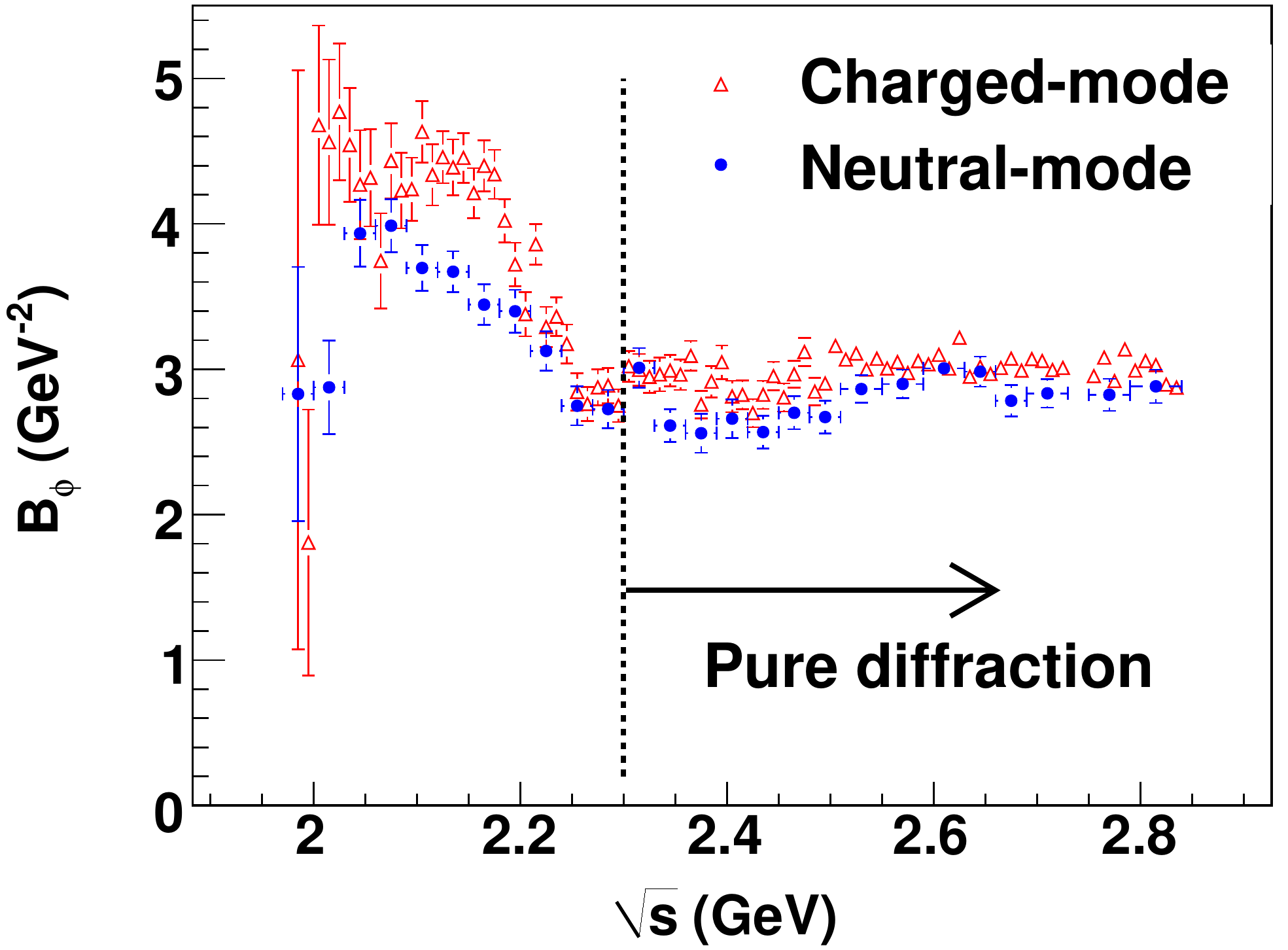} \hspace{-4.5cm} \vspace{-10cm}\rotatebox{45}{\hspace{1cm} PRELIMINARY}
\caption{\label{fig:slope} (Color online) The slope parameter $B_\phi$ extracted from a fit to $d\sigma/dt \propto \exp(-B_\phi |t- t_0|)$.}
\end{figure}

Our kinematic coverage is from near production threshold $\sqrt{s} \sim 1.97$~GeV to 2.84~GeV, with 10-MeV and 30-MeV-wide $\sqrt{s}$ bins for the charged- and the neutral-modes, respectively, and a 0.1 binning in $\cos \theta^\phi_{\mbox{\small c.m.}}$. Fig.~\ref{fig:dcs} shows our preliminary differential cross sections in a forward angle bin for both the charged- (red circles) and neutral-mode (blue triangles). The $\sqrt{s} \sim 2.1$~GeV ``bump'' seen in the previous LEPS data~\cite{mibe} is prominent. It is interesting to note that this structure appears in both the charged- {\em and} the neutral-mode signifying that an interference between the $K^+ \Lambda(1520)$ and $\phi p $ channels, as in the case of the charged-mode~\cite{ozaki}, can not be the predominant cause behind this. It is possible, however, that the two channels couple (rescattering effect) and it does not matter whether the $\phi$ subsequently decays via the charged- or the neutral-mode. We also note here that Ref.~\cite{kiswandhi} ascribes this structure to an $s$-channel resonance. Previous world data for the the $\phi$ is generally very scarce and the few data that exist have wide energy bins and limited statistics. Fig.~\ref{fig:barber} shows a comparison with the 1984 Daresbury data~\cite{barber} at $E_\gamma = 3.3$~GeV. Within the error bars, the agreement is generally fair. Our extracted slope parameter $B_\phi$ from a fit to $d\sigma/dt \propto \exp(-B_\phi |t-t_0|)$ in the range $0.45 \leq \cos \theta^\phi_{\mbox{\small c.m.}} \leq 0.95$, is shown in Fig.~\ref{fig:slope}. Below $\sqrt{s} \sim 2.2$~GeV, the slope deviates from the smooth Pomeron behavior and the charged- and neutral-modes show some differences as well,

\section{Spin Density Matrix Elements}

\begin{figure}
\centering
\hspace{-2.5cm} \includegraphics[angle=90,width=2.2in]{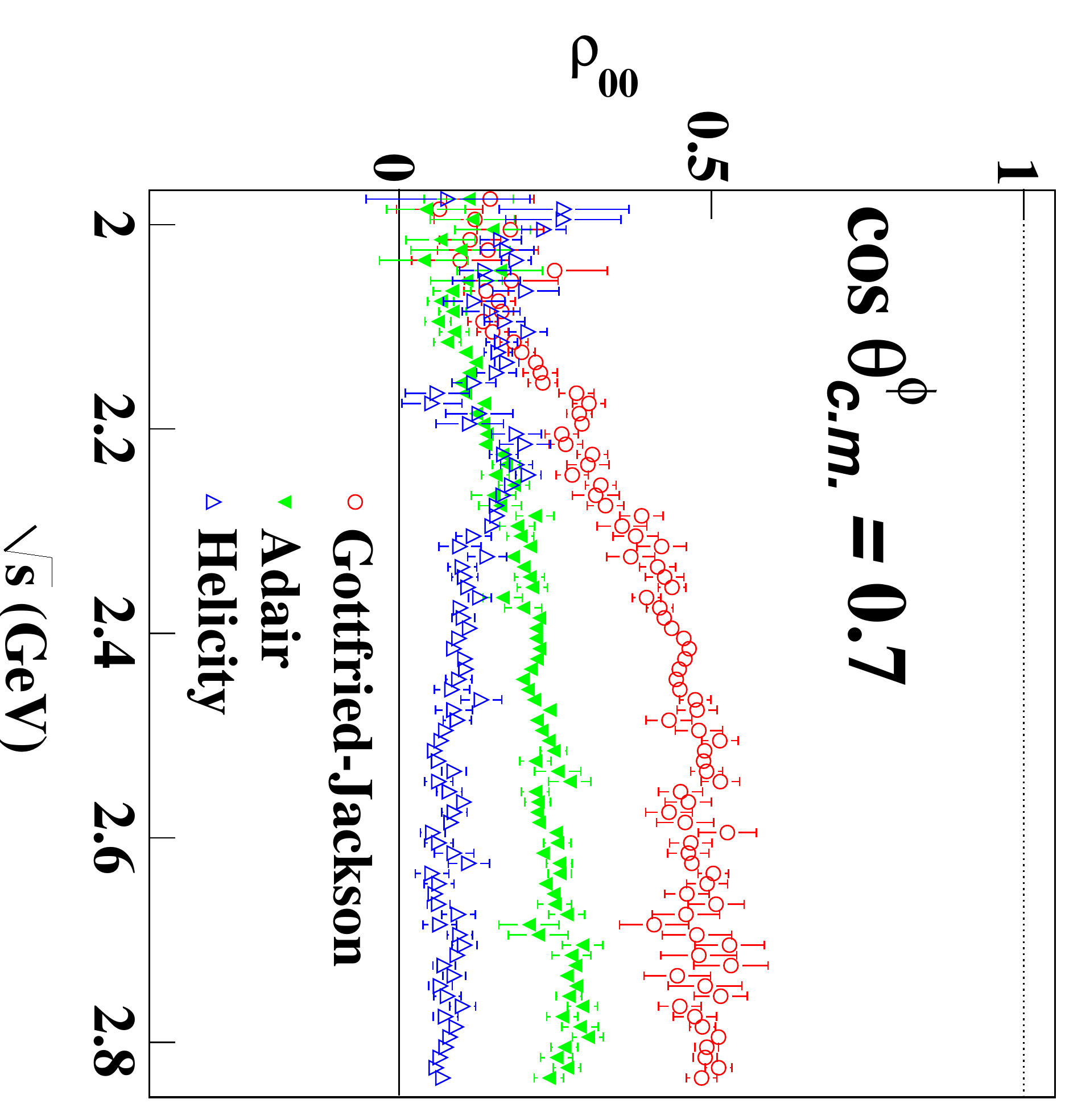} \hspace{-5cm} \vspace{-2cm}\rotatebox{45}{PRELIMINARY}
\caption{\label{fig:sdme} (Color online) Preliminary $\rho^0_{00}$ results for the charged-mode in a forward-angle bin.}
\end{figure}

For a vector meson decaying into spin-0 pseudoscalar mesons, the polarization content is given in terms of four spin density matrices $\rho^i$, $i \in \{0,1,2,3\}$~\cite{schilling}. With an unpolarized beam, as in the current experiment, only the $\rho^0_{MM'}$ elements can be measured. A long puzzle in vector meson photoproduction has been that while the Pomeron exchange process occurs in the $t$-channel, helicity conservation seems to occur in the $s$-channel instead~\cite{gilman}. In the Donnachie-Landshoff (DL) model where the  Pomeron has a $\gamma^{\mu}$ coupling, $s$-channel helicity can be explained somewhat in the limit of massless quarks, but there is no {\em fundamental} reason to expect either $t$- or $s$-channel helicity conservation (TCHC and SCHC, respectively). For SCHC (TCHC), the $\rho^0_{00}$ element has be zero in the Helicity (Gottfried-Jackson) frame, so any deviation from this indicates helicity non-conservation. The recent CLAS $\omega$ data~\cite{williams_omega_prc} shows that both TCHC and SCHC are broken and the present results indicate the same trend for the $\phi$. Fig.~\ref{fig:sdme} shows the charged-mode $\rho^0_{00}$ element in all three frames, Adair, Gottfried-Jackson and Helicity, in a forward-angle bin. Clearly, $\rho^0_{00} \neq 0$ in all three frames. Since the $\phi$ production amplitude is dominated by the Pomeron, this signifies that the DL Pomeron model might have to be revisited to better fit the data.

\bibliographystyle{aipproc}   

\begin{thebibliography}{99}
\bibitem{williams_omega_prc} M.~Williams {\em et al.}, Phys. Rev. C {\bf 80}, 065208 (2009).
\bibitem{cmu_papers} M.~Williams {\em et al.}, Phys. Rev. C \textbf{80}, 045213 (2009); M.~E.~McCracken {\em et al.}, Phys. Rev. C {\bf 81}, 025201 (2010); B.~Dey {\em et al.}, Phys. Rev. C {\bf 82}, 025202 (2010).
\bibitem{mibe} T.~Mibe {\em et al.}, Phys. Rev. Lett. 95, {\bf 182001} (2005).
\bibitem{ozaki} S. Ozaki, A. Hosaka, H. Nagahiro, and O. Scholten, Phys. Rev. C {\bf 80}, 035201 (2009).
\bibitem{kiswandhi} A.~Kiswandhi, J.~J.~Xie, and S.~N.~Yang, Phys. Lett. B {\bf 691}, 214 (2010).
\bibitem{barber} D.~P.~Barber {\em et al.}, Z. Phys. C {\bf 12}, 1 (1982).
\bibitem{schilling} K.~Schilling, P.~Seyboth, and G.~Wolf, Nucl. Phys. B {\bf 15}, 397 (1970).
\bibitem{gilman} F. J. Gilman, J. Pumplin, A. Schwimmer and L. Stodolsky, Phys. Lett. B {\bf 31}, 387 (1970).
\end{thebibliography}

\end{document}